\documentclass[3p,times]{elsarticle}

\usepackage{ecrc}

\volume{00}

\firstpage{1}

\journalname{Nuclear Physics A}

\runauth{F. Gelis}

\jid{nupha}

\jnltitlelogo{Nuclear Physics A}

\usepackage{etex}
\usepackage{bbm}
\usepackage[all]{xy}
\usepackage{graphics}
\usepackage{multirow}
\usepackage{array} 

\usepackage{amsmath}
\usepackage{amsfonts,amsbsy}
\usepackage{amssymb}
\usepackage{bm}
\usepackage{color}
\definecolor{linkcolor}{rgb}{0.4,0.1,0.1}
\definecolor{bibcolor}{rgb}{0.4,0.1,0.1}
\usepackage[colorlinks,linkcolor=linkcolor,citecolor=bibcolor,pdfpagelabels,hyperindex=true]{hyperref}

\biboptions{square,comma,numbers,sort&compress}

\def\empile#1\over#2{\mathrel{\mathop{\kern 0pt#1}\limits_{#2}}}
\def\bs{\boldsymbol}

\def\TODO#1{}

\def\p{{\boldsymbol p}}

\def\k{{\boldsymbol k}}

\def\x{{\boldsymbol x}}

\def\u{{\boldsymbol u}}
\def\v{{\boldsymbol v}}

\newcommand{\slL}{\raise.15ex\hbox{$/$}\kern-.53em\hbox{$L$}}
\newcommand{\slP}{\raise.15ex\hbox{$/$}\kern-.53em\hbox{$P$}}
\newcommand{\slD}{\raise.15ex\hbox{$/$}\kern-.53em\hbox{$D$}}
\newcommand{\slp}{\raise.1ex\hbox{$/$}\kern-.63em\hbox{$p$}}
\newcommand{\slq}{\raise.1ex\hbox{$/$}\kern-.53em\hbox{$q$}}
\newcommand{\slv}{\raise.1ex\hbox{$/$}\kern-.63em\hbox{$v$}}
\newcommand{\slR}{\raise.15ex\hbox{$/$}\kern-.53em\hbox{$R$}}
\newcommand{\slQ}{\raise.15ex\hbox{$/$}\kern-.53em\hbox{$Q$}}
\newcommand{\slK}{\raise.15ex\hbox{$/$}\kern-.53em\hbox{$K$}}
\newcommand{\slk}{\raise.15ex\hbox{$/$}\kern-.53em\hbox{$k$}}
\newcommand{\slSigma}{\raise.15ex\hbox{$/$}\kern-.53em\hbox{$\Sigma$}}
\newcommand{\slcalP}{\raise.15ex\hbox{$/$}\kern-.63em\hbox{$\cal P$}}
\newcommand{\slcalA}{\raise.15ex\hbox{$/$}\kern-.63em\hbox{$\cal A$}}
\newcommand{\slA}{\raise.15ex\hbox{$/$}\kern-.73em\hbox{$A$}}
\newcommand{\slbfA}{\raise.15ex\hbox{$/$}\kern-.73em\hbox{${\imb A}$}}
\newcommand{\slpartial}{\raise.15ex\hbox{$/$}\kern-.53em\hbox{$\partial$}}
\newcommand{\sla}{\raise.15ex\hbox{$/$}\kern-.53em\hbox{$a$}}
\newcommand{\slb}{\raise.15ex\hbox{$/$}\kern-.53em\hbox{$b$}}
\newcommand{\slc}{\raise.15ex\hbox{$/$}\kern-.53em\hbox{$c$}}
\newcommand{\slC}{\raise.15ex\hbox{$/$}\kern-.63em\hbox{$C$}}

\def\colorb{}
\def\colorc{}
\def\colord{}

\begin{document}

\begin{frontmatter}



\title{Initial state in relativistic nuclear collisions\\ and Color Glass Condensate}

%% Single author (and collaboration) - please insert
\author{Fran\c{}cois Gelis}
%\fntext[col1] {A list of members of the XYZ Collaboration and acknowledgements can be found at the end of this issue.}
\address{Institut de Physique Th\'eorique, CEA/Saclay, 91191 Gif sur Yvette cedex, France}

%% For multiple authors, replace the above by:

%\author[label1]{Author1}
%\author[label2]{Author2}

%\address[label1]{Address 1}
%\address[label2]{Address 2}

\begin{abstract}
  In this talk, I discuss recent works related to the pre-hydrodynamical stages of
  ultra-relativistic heavy ion collisions.
\end{abstract}


\end{frontmatter}

% \linenumbers


\section{Introduction}
\label{sec:intro}
Hydrodynamical models are very successful at reproducing bulk
observables in high energy heavy ion collisions. However, it is a long
standing puzzle to understand from the underlying Quantum
Chromodynamics (QCD) why this description is so effective. Indeed, the
description of the early stages of heavy ion collisions which is most
closely related to QCD --the Color Glass Condensate (CGC) framework--
predicts at the very beginning of the fireball evolution a situation
which is very different from a quasi perfect fluid.

\begin{figure}[htbp]
\resizebox*{11cm}{!}{\includegraphics{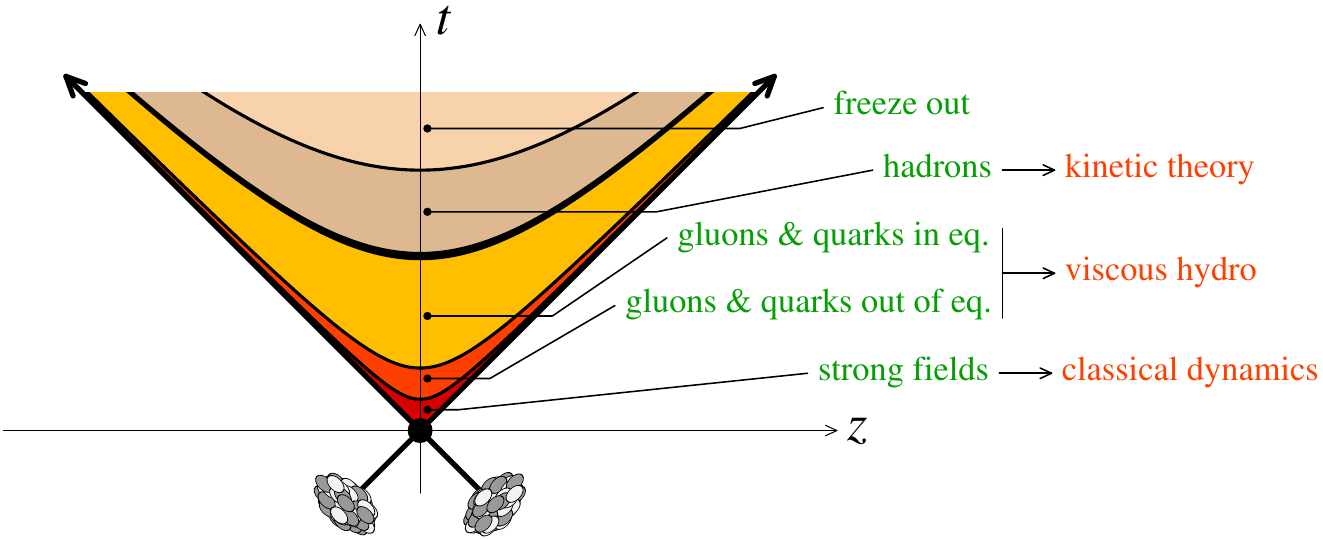}}\hskip 1mm
\resizebox*{5.5cm}{!}{\includegraphics{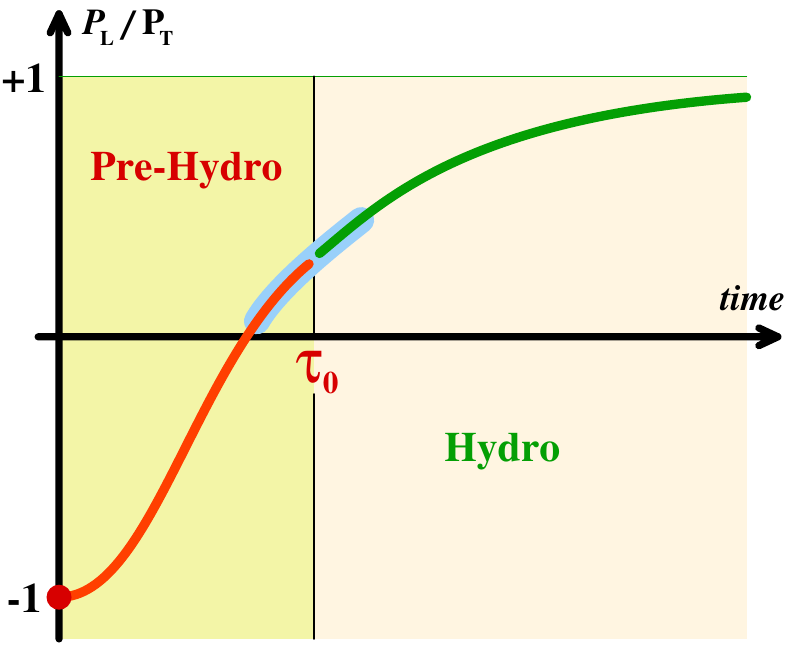}}
\caption{\label{fig:stages}Left: stages of a heavy ion collision. Right: matching to hydrodynamics.}
\end{figure}

The purpose of this talk is to discuss recent works aiming at a first
principles CGC description of the early stages of heavy ion
collisions, where by ``early stages'' we mean the pre-hydrodynamical
evolution (left part of fig.~\ref{fig:stages}). Ultimately, the goal
is to have a description of these early stages that explains how the
hydrodynamical behavior develops and that matches smoothly into
hydrodynamics (right part of fig.~\ref{fig:stages}), in such a way
that the time $\tau_0$ at which the switching happens becomes
unessential (in the same spirit as a factorization scale for parton
distributions).
Note that the CGC, despite being is a weakly coupled framework, can be
the siege of strong interactions because the color fields are large,
of order $g^{-1}$ (or equivalently the gluon occupation number is of
order $g^{-2}$).

\section{Color Glass Condensate}
\label{sec:cgc}
\begin{figure}[htbp]
\resizebox*{6.5cm}{!}{\includegraphics{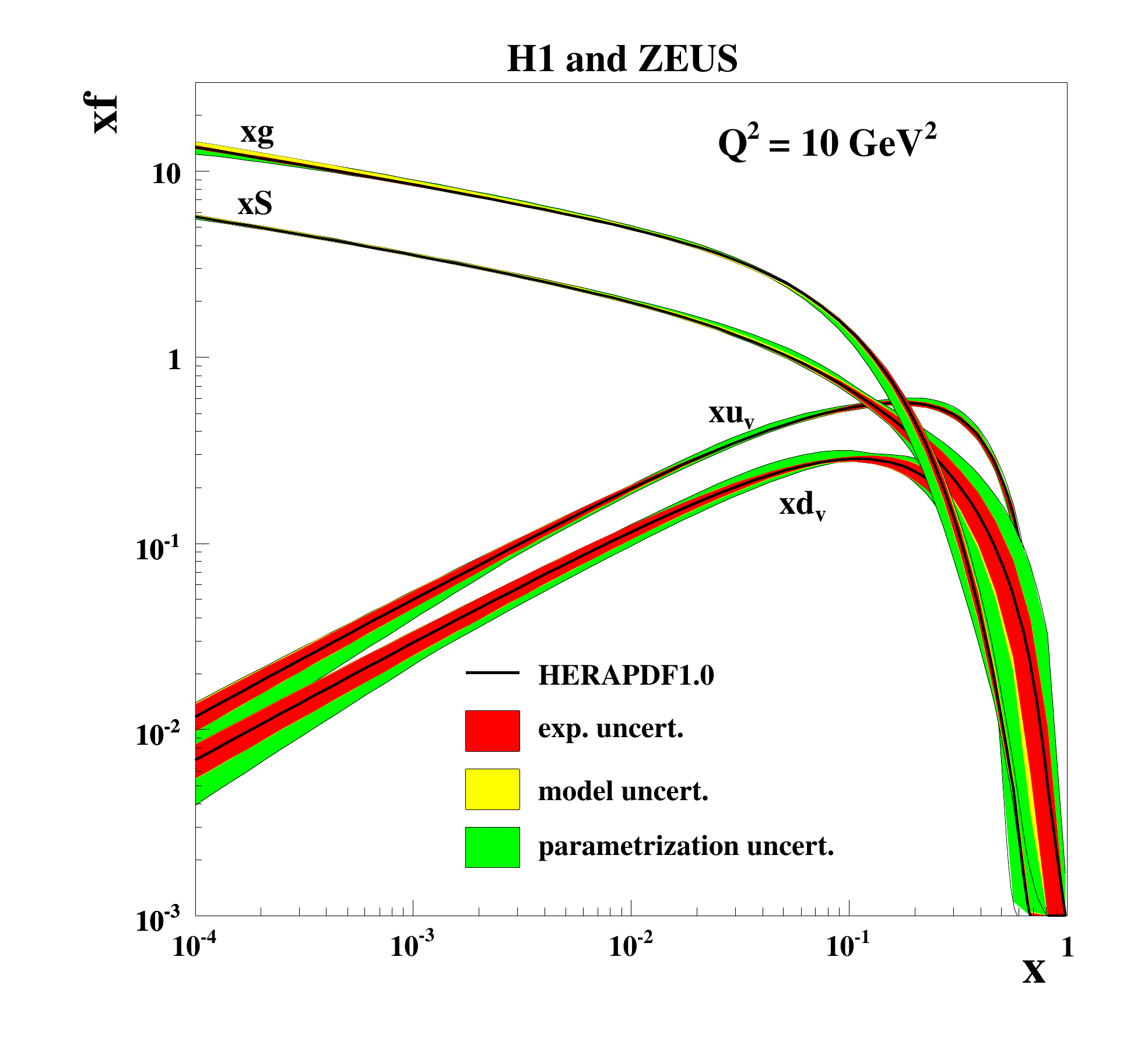}}\hskip 15mm
\raise 17mm\hbox to 7cm{\resizebox*{7cm}{!}{\includegraphics{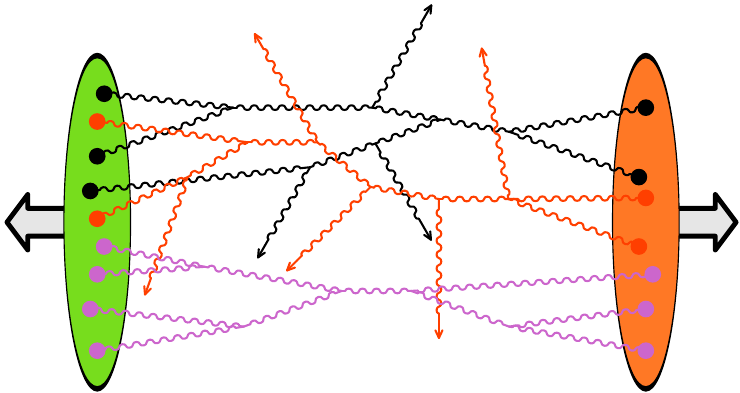}}}
\caption{\label{fig:pdf}Left: parton distributions in a proton. Right:
  multiparton scattering at high gluon density.}
\end{figure}

In high energy heavy ion collisions, most particles are produced with
a comparatively small transverse momentum of a few GeV at most. Given
the longitudinal momentum of the incoming nucleons at the LHC energy,
their constituents are probed with a longitudinal momentum fraction
$x\lesssim 10^{-3}$, where the gluon distribution is very large (left
plot in fig.~\ref{fig:pdf}).
In this regime of large gluon density, multigluon processes become
important, as illustrated in the right panel of the figure
\ref{fig:pdf}. These non-linear effects are one of the manifestations
of {\sl gluon saturation} \cite{GriboLR1,MuellQ1}, which plays a role
for transverse momenta $k_\perp^2\lesssim Q_s^2$, where $Q_s$ is an
$x$-dependent momentum scale known as the saturation momentum (roughly
speaking, $\alpha_s^{-2}Q_s^2$ is the gluon density per unit of
transverse area, and $\alpha_s^{-2}Q_s^2/k_\perp^2$ can be viewed as
the gluon occupation number).
\begin{figure}[htbp]
\resizebox*{6cm}{!}{\includegraphics{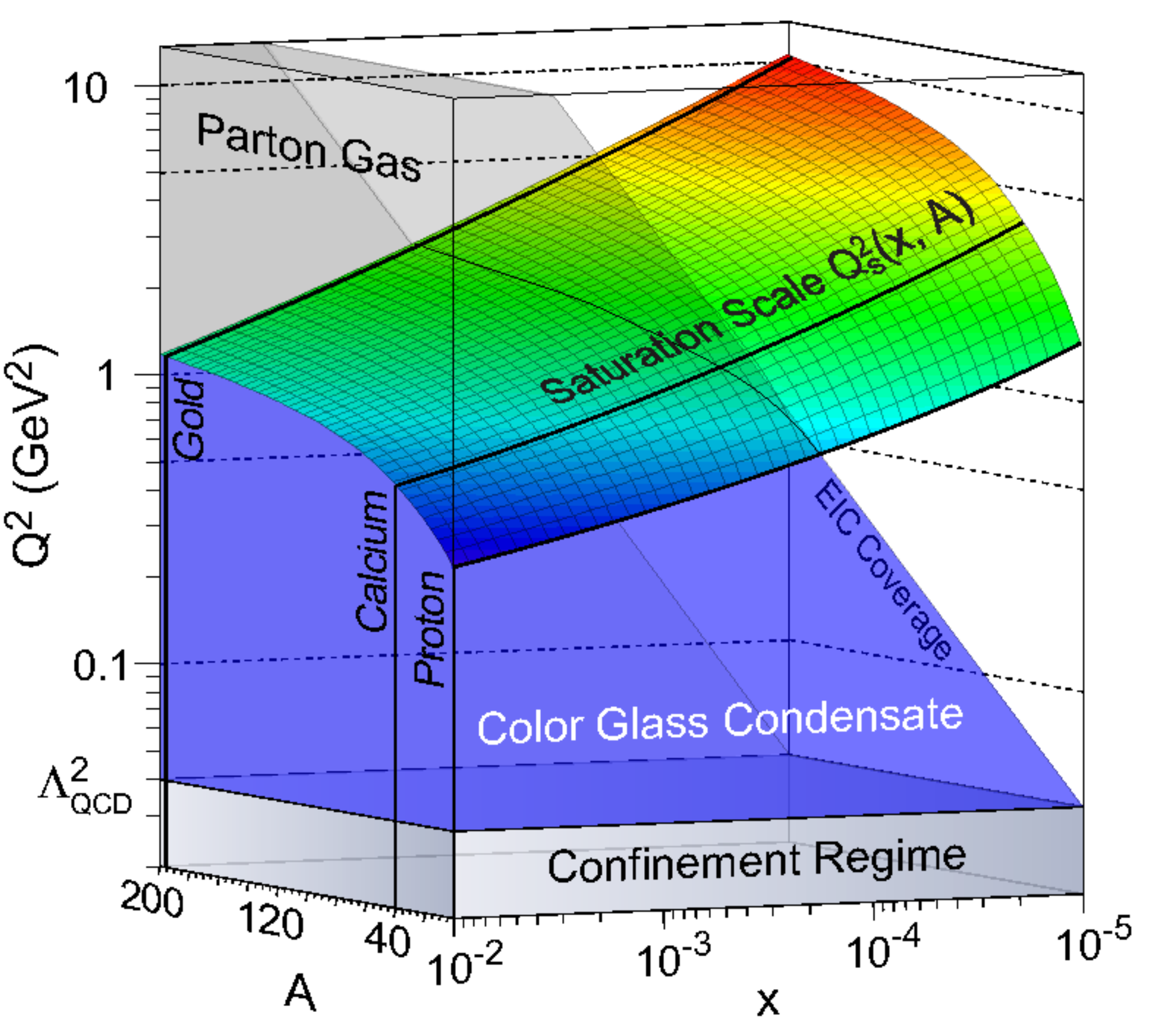}}\hskip 4mm
\raise 2mm\hbox to 10cm{\resizebox*{10cm}{!}{\includegraphics{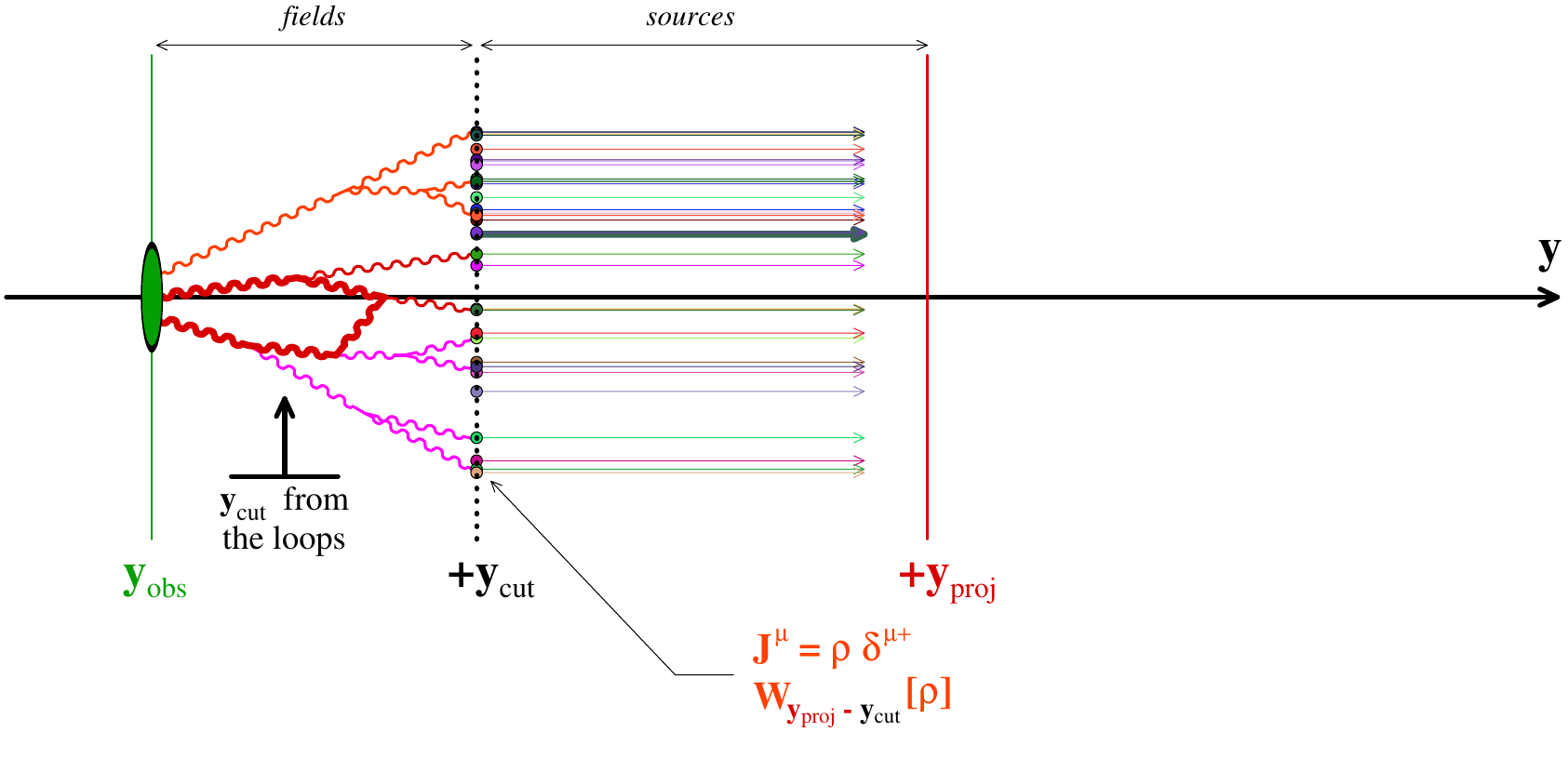}}}
\caption{\label{fig:sat}Left: $x$ and $A$ dependence of $Q_s$ (from
  \cite{DeshpEM1}). Right: high energy factorization in the CGC
  framework.}
\end{figure}
The dependence of $Q_s$ on $x$ and the mass number $A$ is shown in the
left plot of fig.~\ref{fig:sat}.  In the Color Glass Condensate
framework \cite{McLerV1,McLerV2,GelisIJV1}, the partons that have a
longitudinal momentum above a cutoff $\Lambda$ are treated as a color
current $J^\mu=\delta^{\mu+}\rho(\x_\perp)$ along the light-cone,
while those that have a longitudinal momentum below the cutoff (mostly
gluons) are described as usual gauge fields.  The transverse color
distribution of the fast partons, $\rho(\x_\perp)$, fluctuates
event-by-event, and the CGC only provides its probability distribution
$W[\rho]$. The cutoff $\Lambda$ separating the two types of degrees of
freedom is not a physical parameter, and observables should not depend
upon it. This leads to a renormalization group equation for the
distribution $W[\rho]$, known as the JIMWLK equation
\cite{Balit1,IancuLM1,IancuLM2} ($W[\rho]$ must depend on $\Lambda$ in
such a way that it cancels the $\Lambda$ dependence that arises from
loop corrections, as illustrated in the right part of
fig.~\ref{fig:sat} -- one can show that this dependence is universal
for inclusive observables~\cite{GelisLV3,GelisLV4}). A recent
development is the extension of the JIMWLK equation beyond leading
order, with the inclusion of running coupling corrections in
ref.~\cite{LappiM2} and a derivation of the complete NLO result in
refs.~\cite{Grabo1,KovneLM1,KovneLM3}.

In practical calculations, the CGC can be viewed as an effective
Yang-Mills theory coupled to an external color current
$J^\mu_1+J^\mu_2$ (each term corresponding to one projectile). In the
saturation regime, these currents are proportional to $g^{-1}$, and
each order in the $g^2$ expansion is the sum of an infinite set of
graphs. At leading order, observables are obtained from the solution
of the classical Yang-Mills equations $D_\mu
F^{\mu\nu}=J^\nu_1+J^\nu_2$.
\begin{figure}[htbp]
\resizebox*{4.8cm}{!}{\includegraphics{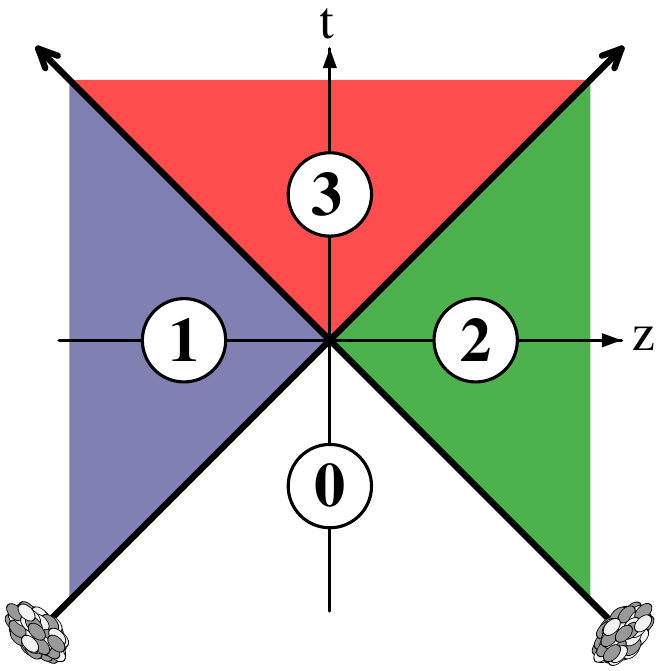}}\hskip 15mm
\raise 4mm\hbox to 7.5cm{\resizebox*{7.5cm}{!}{\includegraphics{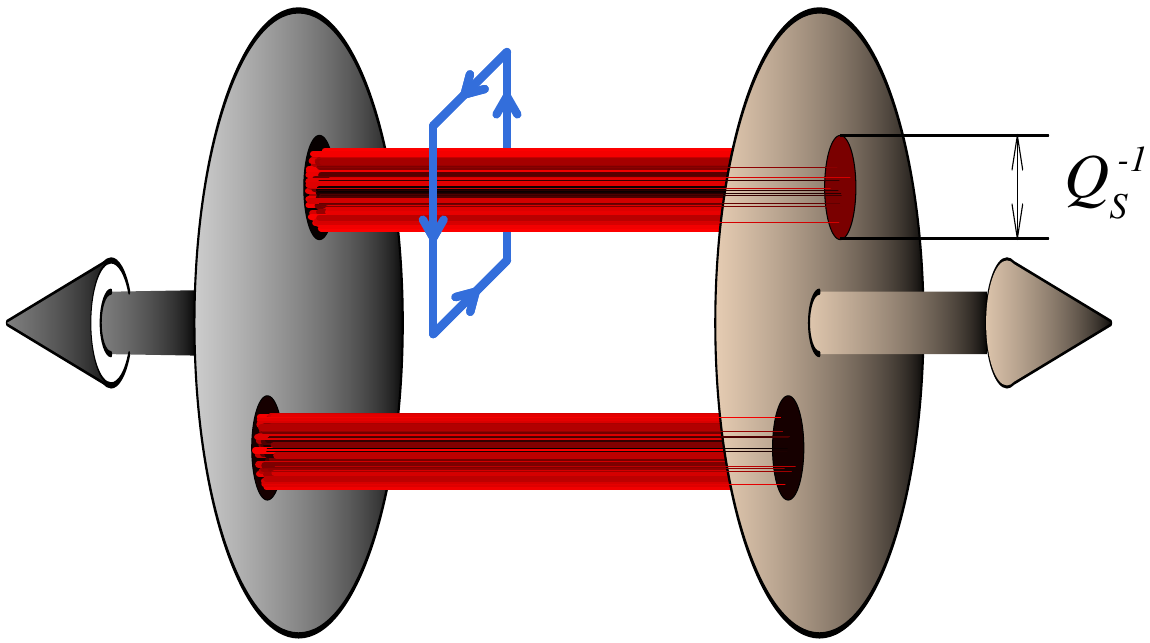}}}
\caption{\label{fig:st}Left: different regions of space-time when
  solving the Yang-Mills equations. Right: structure of the color
  field lines just after the collision.}
\end{figure}
This solution is known analytically in the regions 0,1,2 of the left
plot of fig.~\ref{fig:st}, but must be obtained numerically
\cite{KrasnV1,Lappi1} in the region 3 (it is known analytically at the
interface between 1,2 and 3). Immediately after the collision, the
chromo-electric and chromo-magnetic fields are parallel to the
collision axis \cite{LappiM1} (fig.~\ref{fig:st}, right), forming
longitudinal flux tubes. The typical transverse size of these flux tubes can be
assessed by computing the expectation value of Wilson loops ${\cal W}$
of varying areas: the observation of an approximate area law, ${\cal W}\sim \exp(-{\rm const}\times{\rm
  Area})$, for areas above $Q_s^{-2}$, suggests that the typical
transverse size of the flux tubes is $Q_s^{-1}$~\cite{DumitNP1,DumitLN1}.

\section{Towards hydrodynamical behavior}
\label{sec:flow}
For hydrodynamics to describe the bulk observables in heavy ion
collisions, two conditions must be realized: (i) the ratio of
longitudinal to transverse pressure should not be too small (for the
stability of hydrodynamical codes) and (ii) the shear viscosity to
entropy ratio $\eta/s$ must be small (for an efficient transfer from
spatial to momentum anisotropy, as required by RHIC and LHC data). 

\begin{figure}[htbp]
\resizebox*{7.5cm}{!}{\includegraphics{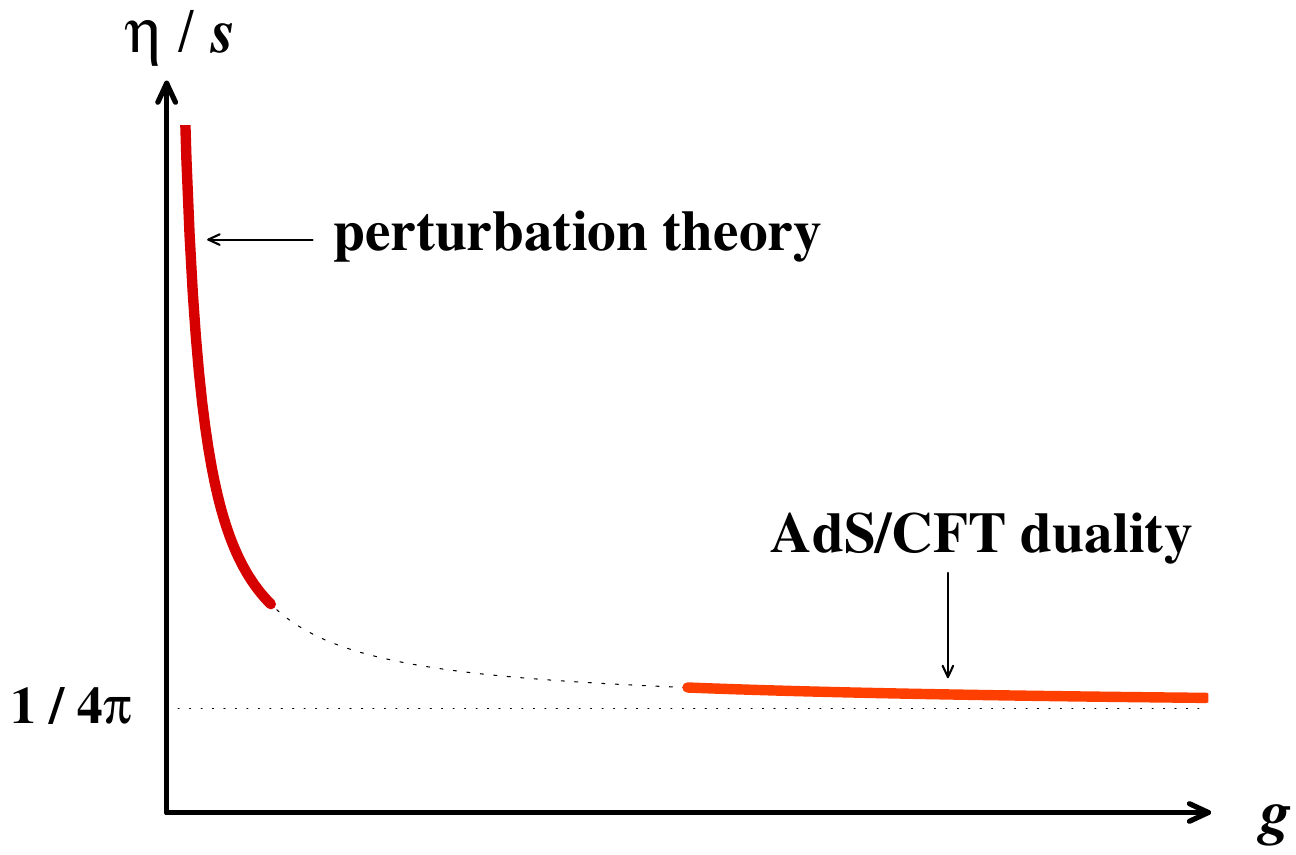}}\hskip 5mm
\raise -5mm\hbox to 8cm{\resizebox*{8cm}{!}{\includegraphics{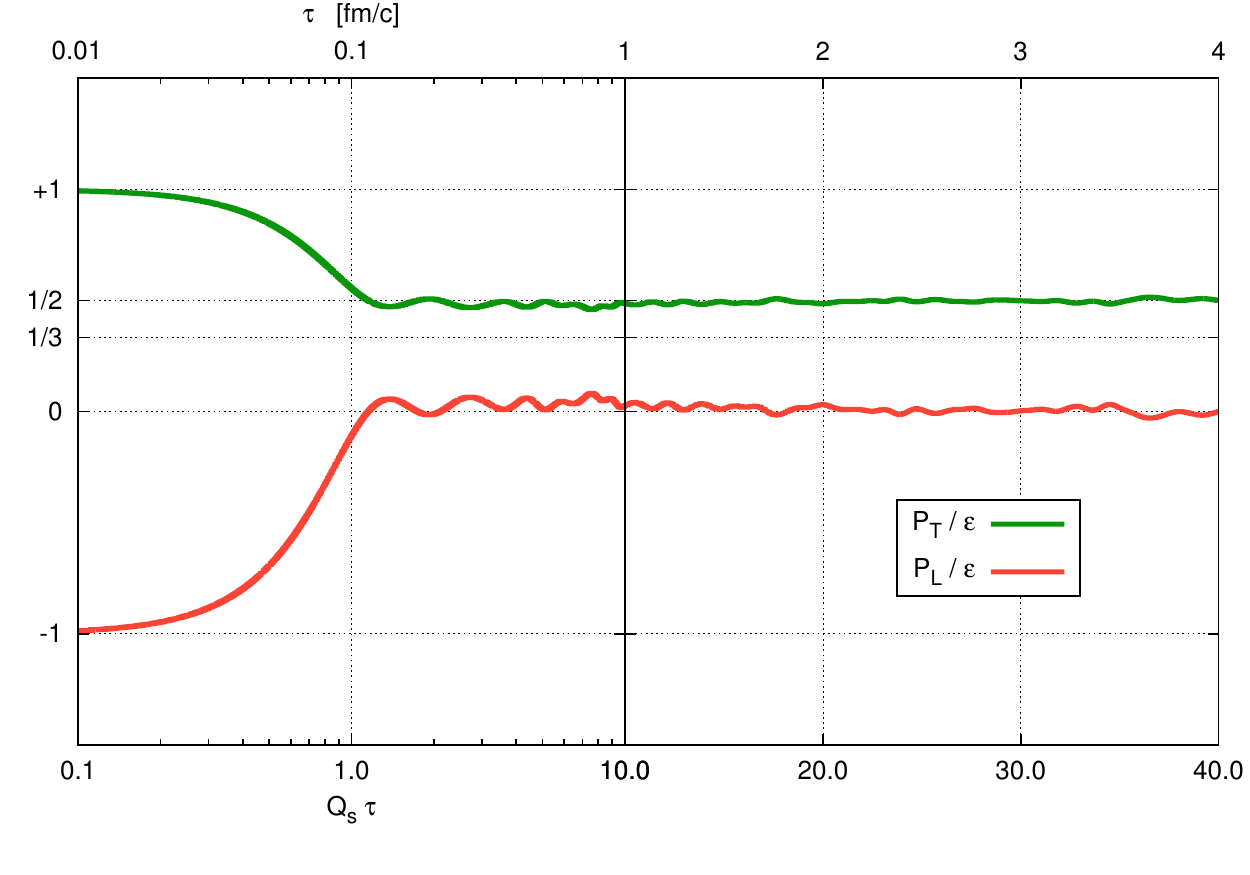}}}
\caption{\label{fig:visco}Left: ratio $\eta/s$ as a function of the
  coupling. Right: $P_{_L}$ and $P_{_T}$ in the CGC at LO.}
\end{figure}

In a weakly interacting system, $\eta/s$ is large. For QCD, it reads
\cite{ArnolMY6} ${\eta}/{s}\approx {5.1}/({{\colorb
    g^4}\ln\left({2.4}/{g}\right)}$) at leading log accuracy. On the
other hand, it has been calculated to be $1/4\pi$ in the strong
coupling limit of SUSY $N=4$ Yang-Mills theories \cite{PolicSS1} (see
the plot on the left of fig.~\ref{fig:visco}). Besides this limit, the
ratio $\eta/s$ can also be small at weak coupling provided that the
occupation number is large. Generically, $\eta/s$ is the ratio of the
mean free path to the De Broglie wavelength of the constituents. In
the CGC, this wavelength is $Q_s^{-1}$, while the inverse mean free
path reads
\begin{equation}
\mbox{(mean free path)}^{-1}\sim \underbrace{\vphantom{\int_\k}g^4 Q^{-2}}_{\mbox{\scriptsize cross section}}\times \underbrace{\int_{\k} \; f_\k}_{\mbox{\scriptsize density}} \underbrace{\vphantom{\int_\k}{\colord (1+f_\k)}}_{{\mbox{\colord\scriptsize Bose}}\atop{\mbox{\colord\scriptsize enhancement}}}\; .
\end{equation}
When $f_\k\sim g^{-2}$ (i.e. in the strong field regime prevalent in
the CGC), the powers of the coupling cancel and one can evade the
conclusion obtained in the weakly interacting scenario.

However, at leading order in the CGC description of heavy ion
collisions, the ratio $P_{_L}/P_{_T}$ has a behavior which is quite
different from the one expected in hydrodynamics: just after the
collision, $P_{_L}$ is exactly opposite to $P_{_T}$ (this is a generic
feature of longitudinal ${\bs E}$ and ${\bs B}$ fields), and then
grows to become mostly positive while remaining much less than $P_{_T}$ at
all times, as shown in the right plot of fig.~\ref{fig:visco}.
\begin{figure}[htbp]
\resizebox*{6cm}{!}{\includegraphics{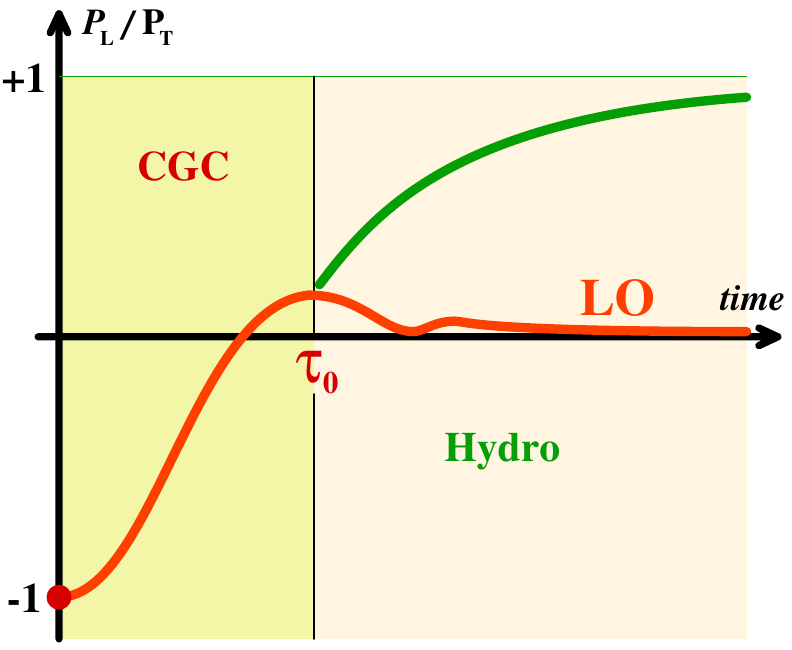}}\hskip 15mm
\raise 0mm\hbox to 6cm{\resizebox*{6cm}{!}{\includegraphics{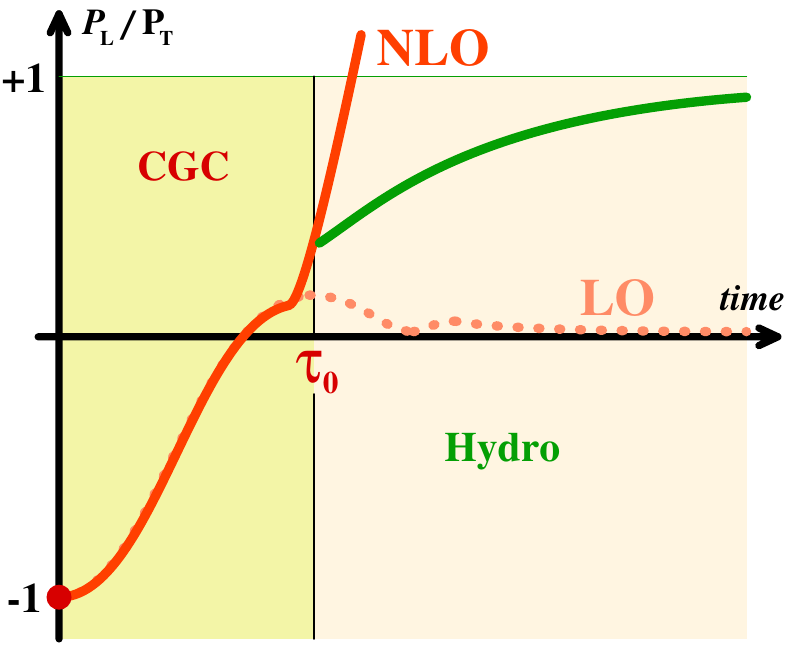}}}
\caption{\label{fig:matching}Matching between the CGC and hydrodynamics (Left: LO, Right: NLO).}
\end{figure}
Therefore, the matching\footnote{To match a CGC calculation to
  hydrodynamics, one should in principle compute the CGC
  energy-momentum tensor, then find its time-like eigenvector to
  determine the local flow velocity and energy density. Then, by
  assuming an equation of state $P=f(\epsilon)$, one get the viscous
  part of the stress tensor as the difference between the full and the
  ideal $T^{\mu\nu}$. Note that very often, this procedure is
  approximated by neglecting the initial flow and the viscous
  stress. } from the CGC at LO to hydrodynamics is not satisfactory
(left part of fig.~\ref{fig:matching}) because the two sides of the
matching describe different physics.

\begin{figure}[htbp]
\resizebox*{6.5cm}{!}{\includegraphics{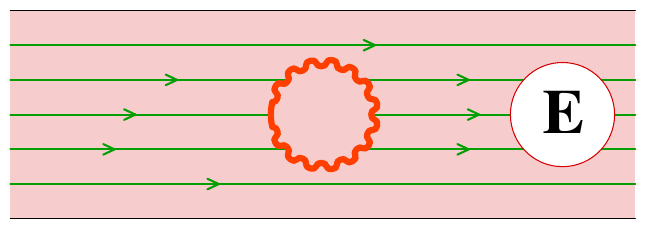}}\hskip 15mm
\raise -17mm\hbox to 6.5cm{\resizebox*{6.5cm}{!}{\includegraphics{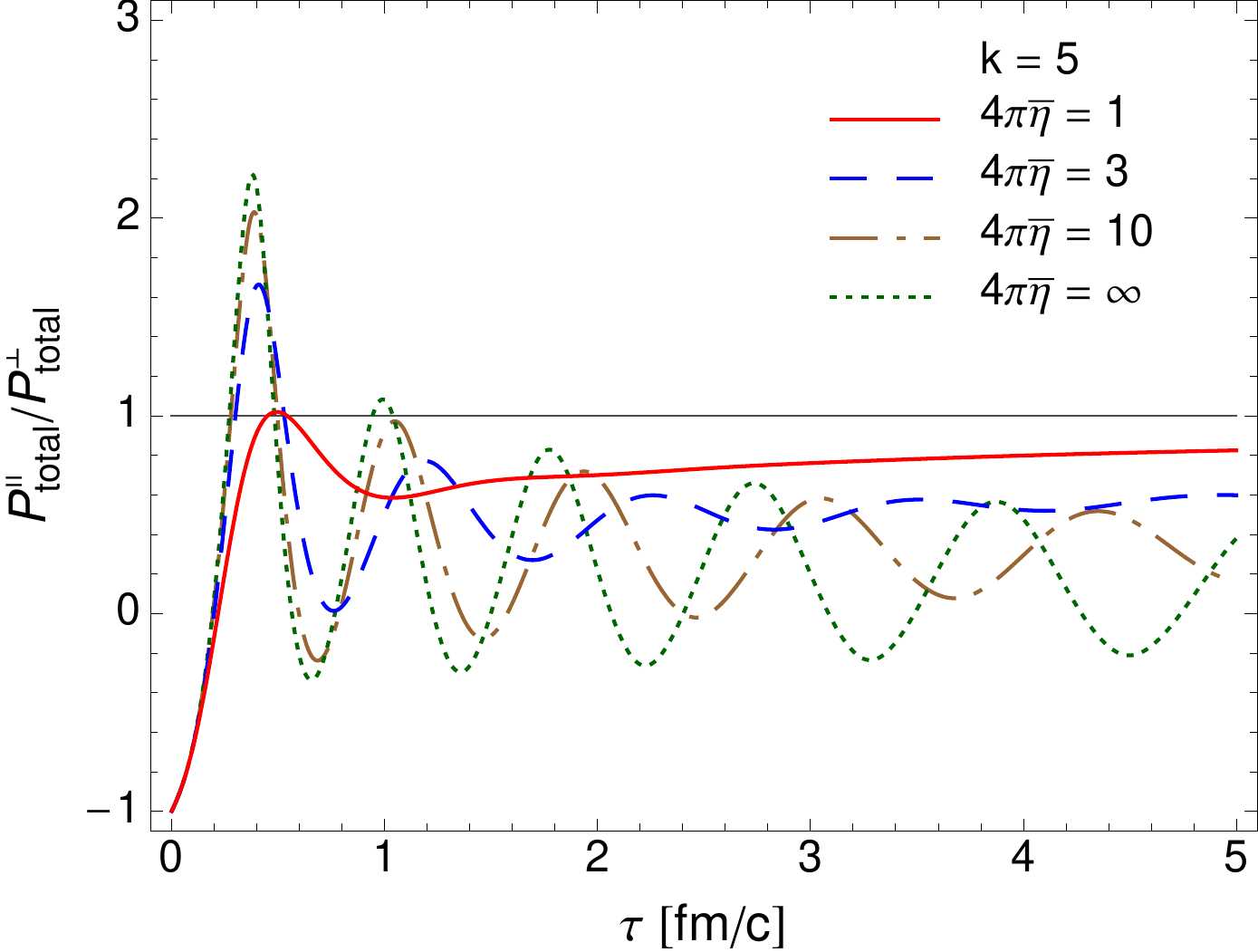}}}
\caption{\label{fig:schwinger}Left: one-loop embedded in a background
  field. Right: ratio $P_{_L}/P_{_T}$ in the color flux tube model of
  ref.~\cite{RybleF1}.}
\end{figure}

The next to leading order CGC result is affected by instabilities in
the classical solutions of the Yang-Mills equations: rapidity
dependent perturbations grow exponentially with time, and induce a
similar behavior in the longitudinal pressure. At NLO, the matching of
the CGC to hydrodynamics is still unnatural, as illustrated in
the right plot of fig.~\ref{fig:matching}. As shown in the left part
of fig.~\ref{fig:schwinger}, the NLO amounts to a one-loop correction
embedded in the classical color fields obtained at LO. This loop has
an imaginary part, related to the possibility of gluon pair production
by the background field. However, in a strict NLO calculation, there
is no feedback from the produced gluons on the background field,
which leads to a runaway behavior. 
This physics has been modeled in a color flux tube model
\cite{RybleF1}, where the produced gluons obey a Boltzmann equation with a
source term reflecting their production by the background field (via
the Schwinger mechanism in this model), and contribute to an induced
color current that feeds back in the Yang-Mills equation that governs
the color fields. In this simple model, it is observed that the ratio
$P_{_L}/P_{_T}$ approaches 1 when the collision rate is tuned so that
$4\pi\eta/s$ is of order one.

\begin{figure}[htbp]
\resizebox*{7.5cm}{!}{\includegraphics{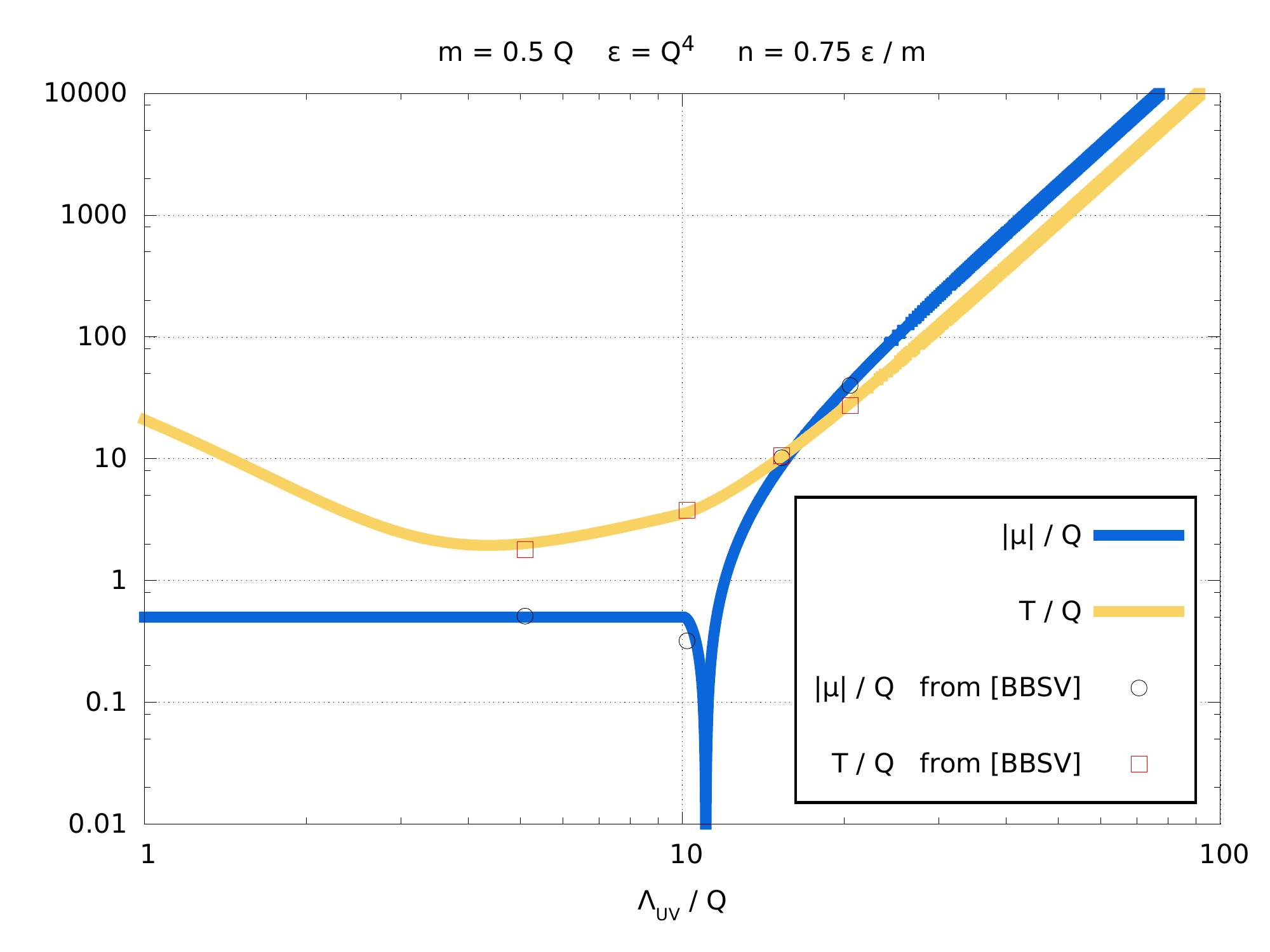}}\hskip 4mm
\raise 10mm\hbox to 8.5cm{\resizebox*{8.5cm}{!}{\includegraphics{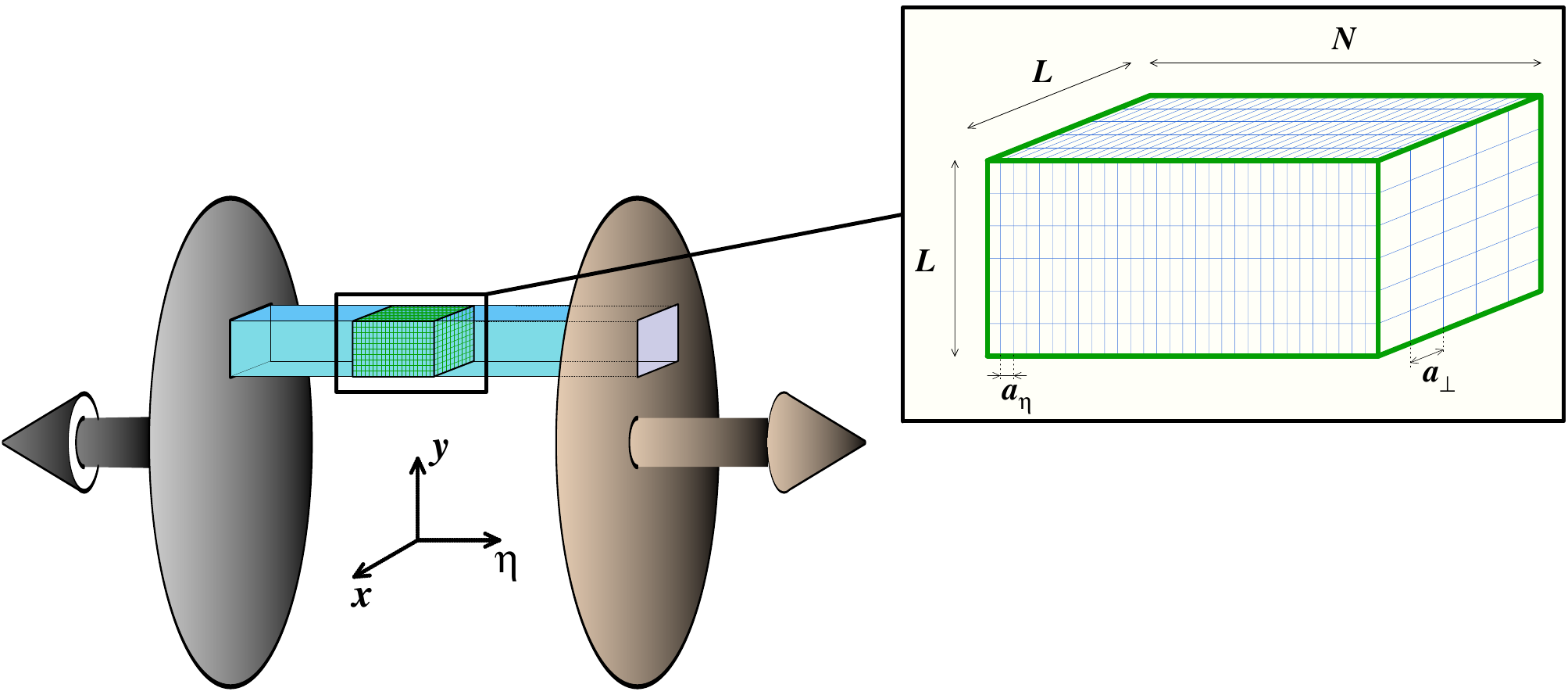}}}
\caption{\label{fig:cutoff}Left: dependence on the UV cutoff of the
  temperature and chemical potential at late times. Right: lattice
  setup for simulating heavy ion collisions in the CSA.}
\end{figure}

In order to study this physics in the CGC framework, one must go
beyond fixed order calculations. Indeed, simple power counting
arguments suggest that the exponential in time growth seen at NLO
becomes even worse at higher fixed loop orders.  Given the central
role played by classical color fields in the CGC, the {\sl classical
  statistical approximation} (CSA) is a natural extension to achieve
such a resummation. In a broad sense, the CSA amounts to performing an
average of classical solutions of the field equations over an ensemble
of initial conditions. Thanks to the fact that one is solving the
fully non-linear equations of motion, the CSA is free of the
exponentially growing terms encountered in fixed order calculations
(in contrast, the NLO calculation amounts to solving the linearized
equations of motion over some background), for any theory where the
Hamiltonian is bounded from below. In studies of heavy ion collisions,
the CSA is easy to implement, by discretizing the $\x_\perp,\eta$
coordinates, as illustrated in the right panel of
fig.~\ref{fig:cutoff} (contrary to the CGC at LO, it is now necessary
to keep the rapidity dependence of the fields because the fluctuations
of the initial conditions break the boost invariance of individual
field configurations -- even though the physics is boost invariant on
average).

An important issue when using the CSA is that of the ultraviolet
divergences. Roughly speaking, the fluctuations of the initial fields
can be categorized in two kinds, vacuum fluctuations and quasiparticle
excitations, that can be easily seen in the bare propagator $G_{22}$ of the
retarded/advanced formalism,
\begin{eqnarray}
&&G_{22}(p)\sim \Big(f_0(\p)+\frac{1}{2}\Big)\,\delta(p^2)\; .\\
&&\mbox{\scriptsize quasiparticles}\hookleftarrow\mbox{\hskip 6mm}
\hookrightarrow\mbox{\scriptsize vacuum fluctuations}\nonumber
\end{eqnarray}
The vacuum fluctuations have a flat spectrum in momentum space, while
the fluctuations associated to quasiparticles have a spectrum that
fall like the initial distribution $f_0(\p)$.  The
quasiparticle-induced fluctuations lead to super-renormalizable
contributions \cite{AartsS1}, provided that $f_0(\p)$ falls at least as
fast as $p^{-1}$. In contrast, the vacuum fluctuations lead to
ultraviolet divergences. Moreover, since the CSA misses some quantum
contributions of the full theory, it is non-renormalizable \cite{EpelbGW1},
as can be seen by the presence of ultraviolet divergences in
self-energies that have no associated operator in the Lagrangian,
\setbox1\hbox to
2.5cm{\resizebox*{2.5cm}{!}{\includegraphics{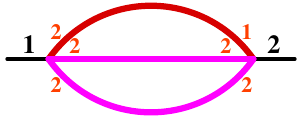}}}
\begin{equation}
\mbox{Im}\,\raise -4.5mm\box1
=
-\frac{g^4}{1024\pi^3}
\;\left({\colorb \Lambda_{_{\rm UV}}^2}-\frac{2}{3}p^2\right)\; .
\end{equation}
When initializing the CSA with vacuum fluctuations, one can for
instance observe a strong dependence on the ultraviolet cutoff of the
state reached by the system at late times, as shown in the plot
\cite{BergeBSV3} on the right of fig.~\ref{fig:cutoff}. It appears that
this sensitivity can be minimized when the UV cutoff is in the range
$3-6$ times the physical scale, but one should keep in mind when using
this type of initial conditions that the lack of renormalizability
prevents a proper continuum limit.

\begin{figure}[htbp]
\resizebox*{4.8cm}{!}{\includegraphics{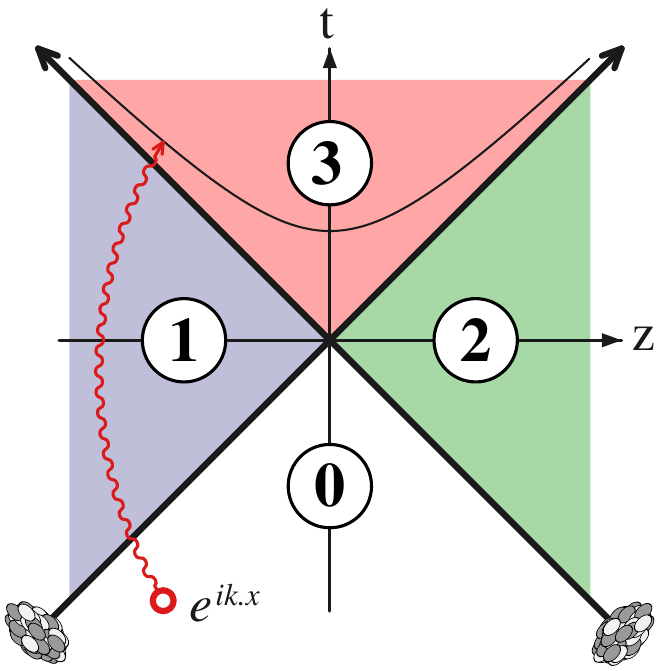}}\hskip 15mm
\raise 0mm\hbox to 6.5cm{\resizebox*{6.5cm}{!}{\includegraphics{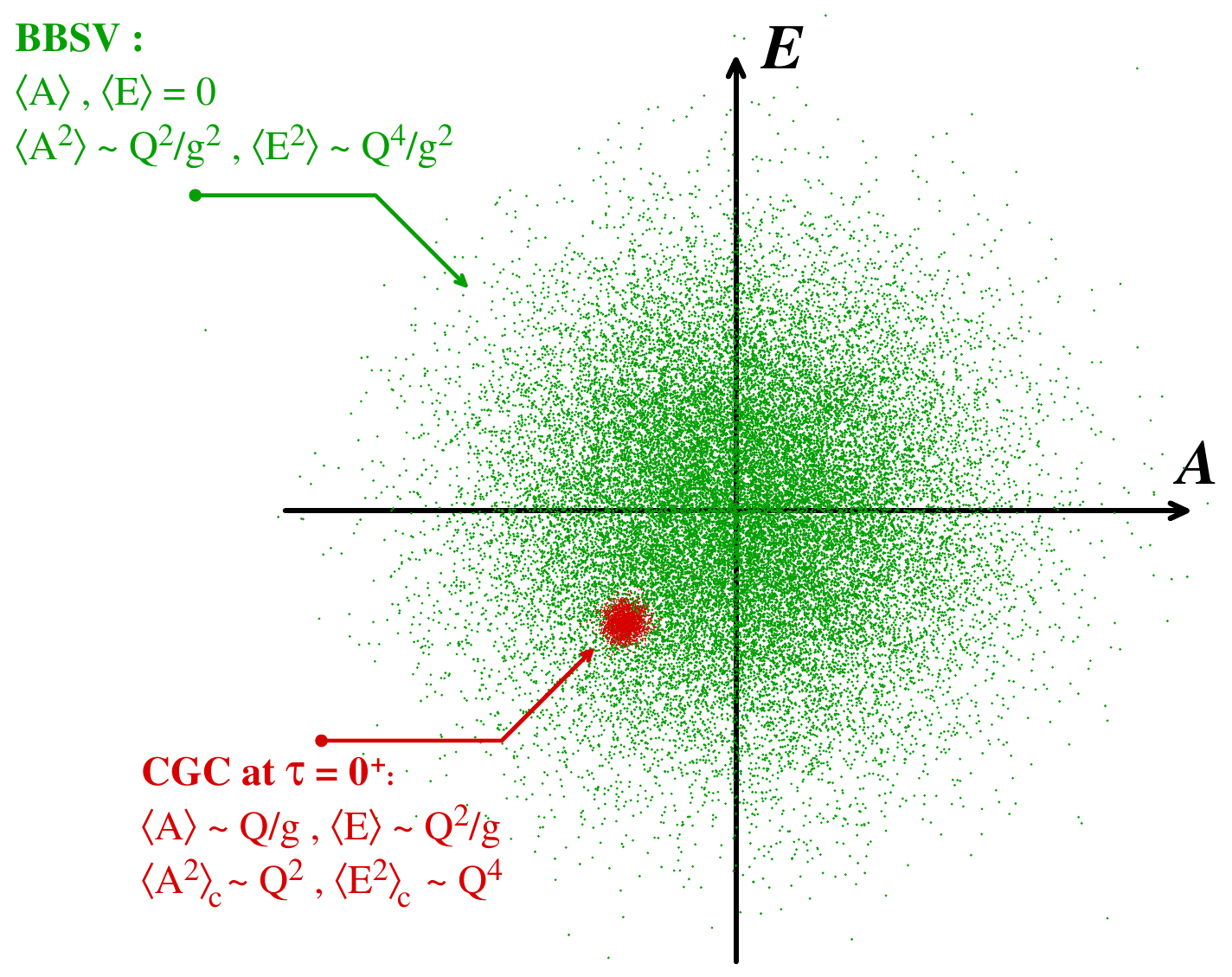}}}
\caption{\label{fig:init}Left: calculation of the initial CGC fluctuations. Right: phase-space distributions in two implementations of the CSA.}
\end{figure}

In the limit where the ensemble of initial conditions narrows down to
a delta function, one recovers the CGC LO result. It can be shown that
there exists a unique Gaussian ensemble of initial conditions such
that the CSA reproduces the exact LO and NLO results, plus a subset of
all higher order terms. This Gaussian ensemble, that can be determined
analytically at $Q_s\tau\ll 1$ \cite{EpelbG2}, is centered on the LO
classical field, and its variance is obtained by solving the
linearized Yang-Mills equations over the LO background, with a plane
wave initial condition in the remote past (left of
fig.~\ref{fig:init})~:
\begin{eqnarray}
\big<{\cal A}^\mu\big>={\cal A}_{_{\rm LO}}^\mu\qquad\!\!\mbox{Variance}
 = \!\!\!\!\int\limits_{{\rm modes\ }\k}\!\!\!\!
{\colorc\frac{1}{2}}\;{\colorb a_\k(\u) a_\k^*(\v)}\qquad
\Big[{\cal D}_\rho {\cal D}^\rho \delta^\nu_\mu - {\cal D}_\mu {\cal D}^\nu
+ig\, {\cal F}_\mu{}^\nu\Big] {\colorb a_\k^\mu}=0\qquad
{\colord\lim_{x^0\to-\infty}} {\colorb a_\k(x)} = e^{ik\cdot x}
\label{eq:init1}
\end{eqnarray}
However, these initial conditions are (dressed)
vacuum fluctuations, and therefore lead to a non-renormalizable CSA.
Other works \cite{BergeBSV1,BergeBSV2} have considered a particle-like Gaussian
ensemble of initial fluctuations, with a vanishing central value and
large fluctuations~:
\begin{equation}
\big<{\cal A}^\mu\big>=0\qquad
  \mbox{Variance}
 = \!\!\!\!\int\limits_{{\rm modes\ }\k} \!\!\!\!{\colorc f_0(\k)}\;
{\colorb a_\k(\u) a_\k^*(\v)}\qquad {\colorb a_\k(x)} \equiv e^{ik\cdot x}
\qquad
{\colorc f_0(\k)}\sim g^{-2}\times\theta(Q_s-k)
\label{eq:init2}
\end{equation}
These initial conditions lead to a proper UV limit, but whether they
can be connected with CGC fields at $Q_s\tau\ll 1$ is unclear at the moment.
In this model, one can play with the gluon distribution $f_0(\k)$ in
order to control the magnitude of the initial occupancy and its anisotropy in
momentum space. The difference between the initial conditions of type
(\ref{eq:init1}) and (\ref{eq:init2}) is illustrated in the plot on
the right of fig.~\ref{fig:init}.

\begin{figure}[htbp]
\resizebox*{8cm}{!}{\includegraphics{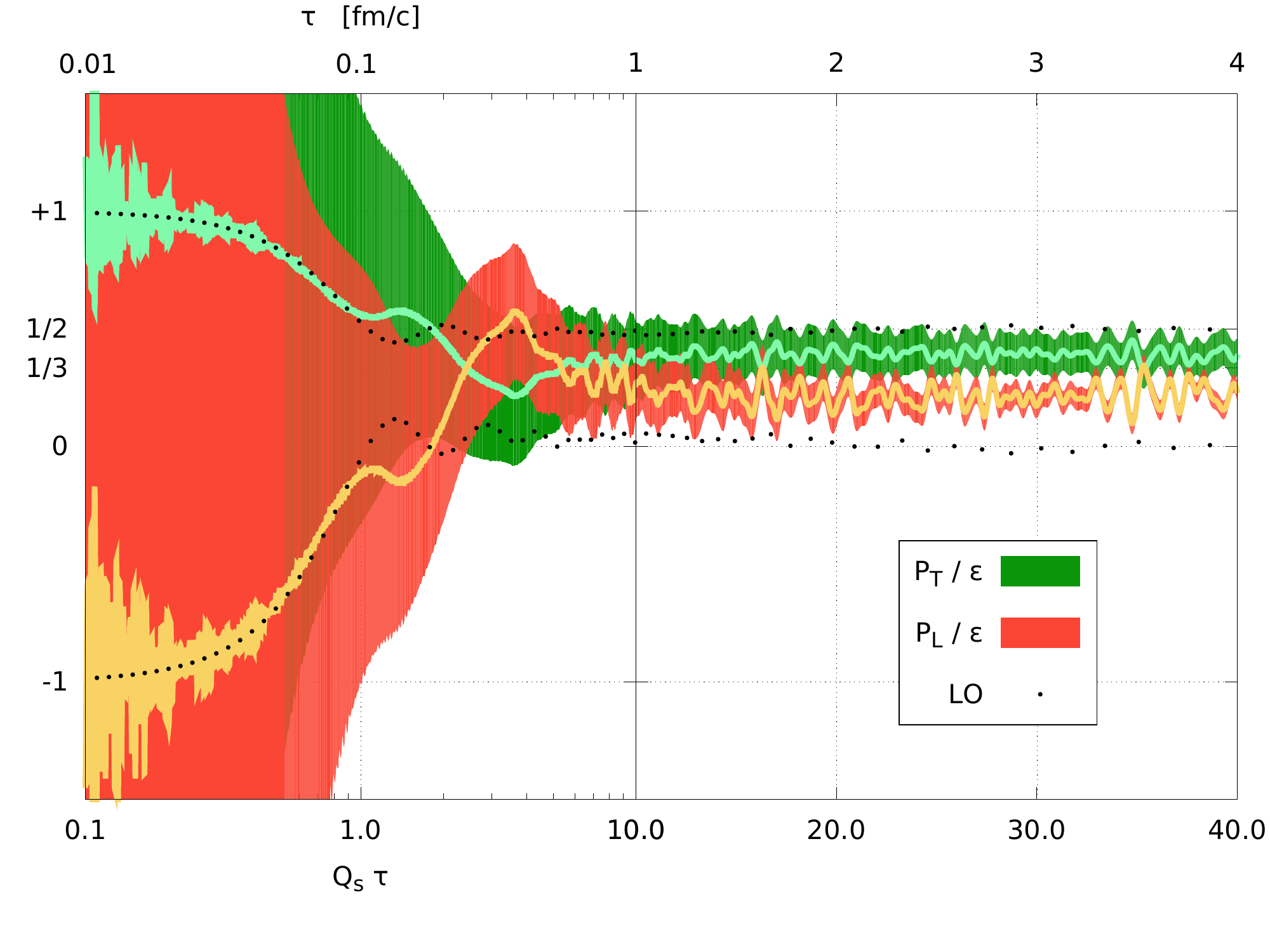}}\hskip 5mm
\raise 2mm\hbox to 8cm{\resizebox*{8cm}{!}{\includegraphics{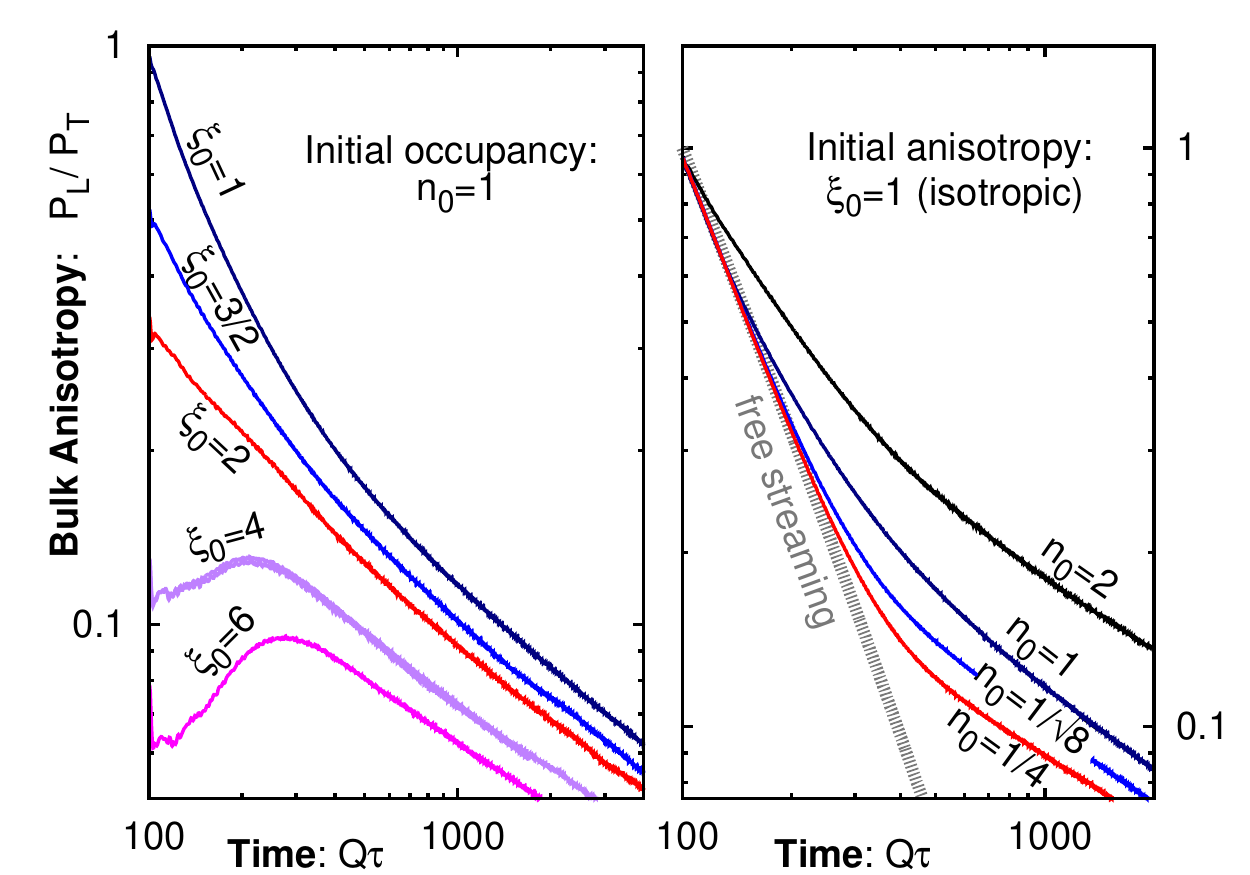}}}
\caption{\label{fig:PLPT}Results of the CSA for the ratio
  $P_{_L}/P_{_T}$. Left~: CGC vacuum-like initial fluctuations at
  $Q_s\tau\ll1$ \cite{EpelbG3}. Right~: particle-like initial fluctuations at
  $Q_s\tau\gg 1$ \cite{BergeBSV1,BergeBSV2}.}
\end{figure}

Some results obtained with these two types of initial conditions are
shown in the figure \ref{fig:PLPT} (Left: CGC initial conditions at
$Q_s\tau\ll 1$, Right: particle-like initial conditions at $Q_s\tau\gg
1$).
With CGC vacuum-like fluctuations, the ratio $P_{_L}/P_{_T}$ is found
to increase significantly above the LO values, for a coupling
$g=0.5$ \cite{EpelbG3}. However, one should keep in mind that this calculation was
performed on a small lattice ($64\times64\times128$) and that because
the initial conditions are vacuum fluctuations it has no continuum
limit.  With particle-like initial fluctuations, and starting at a
much larger time, the study of ref.~\cite{BergeBSV1,BergeBSV2} found a ratio
$P_{_L}/P_{_T}$ that reaches a universal behavior for varying initial
occupations and anisotropies, following a self-similar evolution
\begin{equation}
      f(t,p_\perp,p_z)\sim {\colorb \tau^{-2/3}}\;f_{_S}({\colorb \tau^0} p_\perp,{\colorb \tau^{1/3}}p_z)
      \qquad \frac{P_{_L}}{P_{_T}}\sim \tau^{-2/3}
\end{equation}
intermediate between free streaming,
\begin{equation}
      f(t,p_\perp,p_z)\sim {\colorb \tau^{0}}\;f_{_S}({\colorb \tau^0} p_\perp,{\colorb \tau^{1}}p_z)
      \qquad \frac{P_{_L}}{P_{_T}}\sim \tau^{-2}\;, 
\end{equation}
      and expansion at constant anisotropy~:
\begin{equation}
      f(t,p_\perp,p_z)\sim {\colorb \tau^{\frac{2(\delta-1)}{2+\delta}}}\;f_{_S}({\colorb \tau^{\frac{\delta}{2+\delta}}} p_\perp,{\colorb \tau^{\frac{\delta}{2+\delta}}}p_z)
      \qquad \frac{P_{_L}}{P_{_T}}=\delta\; .
\end{equation}
With this kind of initial conditions, the coupling constant $g$
cancels out and does not affect the time dependence of
$P_{_L}/P_{_T}$. The coupling only affects the time at which the
fields (that decrease with time) become too small for the classical
approximation to be trusted, since they start with a magnitude $\sim
g^{-1}$.  When this happens, one should switch to a description that
has quantum effects built in. In the case of non-expanding systems,
classical statistical simulations have been combined with an effective
kinetic theory description that takes over the evolution of the system
at low occupation numbers, in order to follow the time evolution of
the system from a highly occupied initial condition all the way to thermal
equilibrium \cite{KurkeM3,YorkKLM1,KurkeL1}.

Let us finish this overview by mentioning some works on Bose-Einstein
condensation (BEC), that has been speculated to occur in heavy ion collisions
due to the initially overoccupied gluon distribution. The starting
observation is that in the CGC, the dimensionless ratio
$n/\epsilon^{3/4}$ ($n$ being the number of gluons per unit volume) is
initially $g^{-1/2}$ and is thus large at weak coupling, while it is
of order unity in thermal equilibrium \cite{BlaizGLMV1}. To resolve this
discrepancy, the system can eliminate the excess of gluons via
inelastic processes (e.g. $3\to 2$), or by forming a condensate at
$k=0$ (at asymptotic times, this BEC should disappear because the
gluon number is not conserved). The formation of such a condensate has
been observed in scalar theories studied in the classical statistical
approximation \cite{EpelbG1,BergeS4}. In QCD, kinetic theory computations
do see the formation of a BEC \cite{BlaizLM2}, even if only as a transient
phenomenon when number changing processes are included, while other
calculations show no sign of it \cite{KurkeM3,YorkKLM1}. Even if such a
condensate would form, its phenomenological consequences are unclear
at the moment.  Rather intriguingly, fits of pion spectra at very low
momentum \cite{BegunFR1} tend to support a positive chemical
potential almost equal to the pion mass, which is the value that would
be realized if a transient pion condensate was formed.

\section{Conclusions}
The CGC provides a QCD-based theoretical framework for studying from
first principles the initial stages of heavy ion collisions. The
parton content of the colliding projectiles is represented in the form
of probability distributions $W[\rho]$ for the color charge transverse
density. These distributions obey a universal RG equation --the JIMWLK
equation-- which is now known up to NLO, and enter into factorized
expressions for all inclusive observables.

At early times after the collision of two heavy ions, the CGC can be
viewed as a weakly coupled (because the relevant momentum scale is the
saturation momentum $Q_s\gg\Lambda_{_{\rm QCD}}$, that increases with the collision energy),
but strongly interacting system (because it contains color fields that
are of order $g^{-1}$).

Fixed order calculations do not match properly onto the expected
hydrodynamical behavior. At LO, the longitudinal pressure never
becomes comparable to the transverse one, while at NLO instabilities
make it increase indefinitely in an unphysical way. One can go beyond
this by using the classical statistical method, where one solves
classical field equations with fluctuating initial conditions. At the
moment, two recent works have implemented it with two different types
of initial conditions, leading to different results regarding the
isotropization of the pressure tensor.

\vglue 3mm
\noindent{\bf Acknowledgements~:}
This work is supported by the Agence Nationale de la Recherche project
11-BS04-015-01.


%\bibliographystyle{elsarticle-num}
%\bibliography{biblio}



\end{document}